\documentclass[10pt,a4paper,twoside]{article}
\usepackage{epsfig}
\usepackage{baltlat6}
\usepackage{array}
\usepackage{here}
\usepackage{hyperref}   
\usepackage{changes}
\def\aplt{\ {\raise-.5ex\hbox{$\buildrel<\over\sim$}}\ }

\begin{document}
\vspace{0.5mm}
\setcounter{page}{1}

\titlehead{Baltic Astronomy, vol.\,??, ???--???, 2016}

\titleb{VISUAL BINARY STARS: DATA TO INVESTIGATE\\ FORMATION OF BINARIES}

\begin{authorl}
\authorb{D. Kovaleva,}{}
\authorb{O. Malkov,}{}
\authorb{L. Yungelson,}{}
\authorb{D. Chulkov}{}
\end{authorl}

\begin{addressl}
\addressb{}{Institute of Astronomy, Russian Acad. Sci.,\\
48 Pyatnitskaya St., 119017 Moscow, Russia;
dana@inasan.ru}
\end{addressl}

\submitb{Received: 2016 ????? ?; accepted: 20?? ???? ?}

\begin{summary}

Statistics of orbital parameters of binary stars as well as
statistics of their physical characteristics bear traces of star
formation history. However, statistical investigations of binaries
are complicated by lacking or incomplete observational data and by
a number of observational selection effects.

Visual binaries are the most numerous observed binaries, the
number of pairs exceeds 130000. The most complete list of
presently known visual binary stars was compiled by cross-matching
objects and combining data of the three largest catalogues of
visual binaries. It was supplemented by the data on parallaxes,
multicolor photometry, spectral characteristics of the stars of
this list taken from other catalogues. This allowed us to
compensate partly for the lack of observational data for these
objects. Combined data allowed us to check validity of
observational values and to investigate statistics of the orbital
and physical parameters of visual binaries. Corrections for
incompleteness of observational data are discussed. Obtained
datasets and modern distributions of binary parameters will be
used to reconstruct the initial distributions and parameters of
the function of star formation for binary systems.\end{summary}


\resthead{VISUAL BINARY STARS: DATA}
{D.~Kovaleva, O.~Malkov, L.~Yungelson, D.~Chulkov}

\sectionb{1}{INTRODUCTION}

Binary stars are a significant component of stellar population
that can be observed via a number of methods giving a variety of
data. Statistics of parameters of binary systems first became
subject of interest back in 1920--1930s (\"Opik 1924, Kuiper 1933,
Aitken 1935, Ambartsumian 1937, etc.); it was repeatedly revisited
in 1970--1980s when new challenging results had been obtained, and
remains a matter of discussion currently, since the new methods
and observational techniques allow us to register more binaries:
distant binaries, binaries with fainter components, closer located
components, more remote components (Abt \& Levy 1976, Vereshchagin
et al. 1988, Kraicheva et al. 1989, Duquennoy \& Mayor 1991,
Poveda et al. 1994, 2007, Patience et al. 2002, Abt 2006,
Kouwenhoven et al. 2008, Rahgavan et al. 2010,  Duch\^ene \& Kraus
2013, Tokovinin 2014, Tokovinin \& Kiyaeva 2016, etc.).

It is accepted that the statistics of orbital parameters of binary
stars as well as the statistics of their physical characteristics
may bear traces of star formation history (see, for instance,
Poveda et al. 2007, Tokovinin \& Kiyaeva 2016). These traces are,
however, substantially distorted by evolution and by the effects
of observational selection. The components of so-called wide pairs
do not affect each other in the course of physical evolution, and
orbital parameters of such pairs evidently remain almost unchanged
(the unique active agent is mass, while angular momentum loss via
stellar wind is significant for giants only). This makes
investigation of distributions of wide pairs over various
characteristics a rather attractive task.

In the modern investigations of statistics of binaries, one may
observe three different approaches and, sometimes, their
combination. Namely, some authors consider datasets declared to be
volume-limited; some choose uniform datasets like those in
clusters and associations; some simulate observations based on
theoretical considerations and compare the result to observational
data. Every approach has its advantages and its flaws, and
obviously the complete picture may come as the result of their
combination. In this work, we present datasets compiled to be
modelled by population synthesis methods to solve the inverse
problem of reconstruction of initial parameter distributions of
binary stars.

The majority of wide pairs are observed as visual binaries. The
number of known catalogued visual pairs (those that can be
resolved using a telescope) exceeds 130000. Recently, a new
comprehensive set of data on visual binaries named WCT was
compiled by Isaeva et al. (2015) by cross-matching data from the
current version of The Washington Visual Double Star Catalog (WDS,
Mason et al. 2014), the Catalog of Components of Double \&
Multiple stars (CCDM, Dommanget \& Nys 2002), and the Tycho Double
Star Catalogue (TDSC, Fabricius et al. 2002). Additionally, the
WCT contains parallaxes for more than 14000 pairs, mainly from
Hipparcos data (and the rest from SIMBAD). The WCT catalogue
(together with WDS, CCDM and TDSC) was uploaded into the Binary
stars database, BDB (Kaygorodov et al. 2012, Kovaleva et al.
2015a). A statistical investigation of the WCT data was described
by Kovaleva et al. (2015b). In this paper, we present solution of
some problems we encountered, as well as add and analyze data to
complete construction of the set of data for modelling.

In Section 2, we discuss what can be said about the distribution
over periods for wide pairs based on visual binaries data. In
Section~3, the refinement of spectral classification process and
results are described. In Section~4, we discuss obtaining
distributions over physical parameters of the components. In
Section~5, the results of investigation of the visual binary star
data are summarized.

\begin{figure}[!tH]
\vbox{
\centerline{\psfig{figure=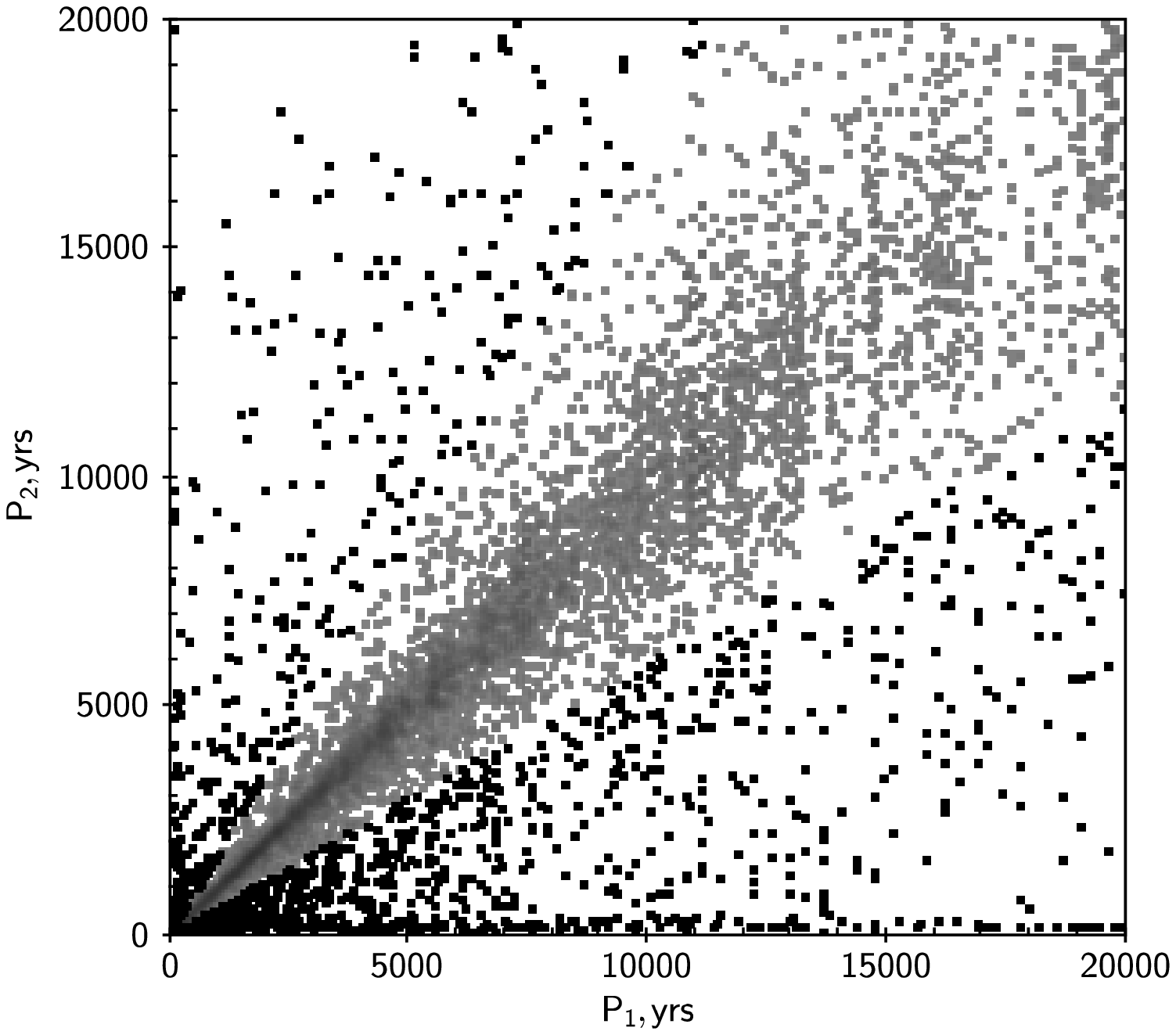,width=100mm,angle=0,clip=}}}
\vspace{1mm} \captionb{1} {Period estimates for the same wide
binaries obtained by different couples of positions (based on data
for 8000+ binaries with three and more positions). Grey dots are
within the sector where  $P_2=P_1\pm 50\% $; this sector contains
$80\%$ of all points.  }
\end{figure}

\begin{figure}[!tH]
\vbox{
\centerline{\psfig{figure=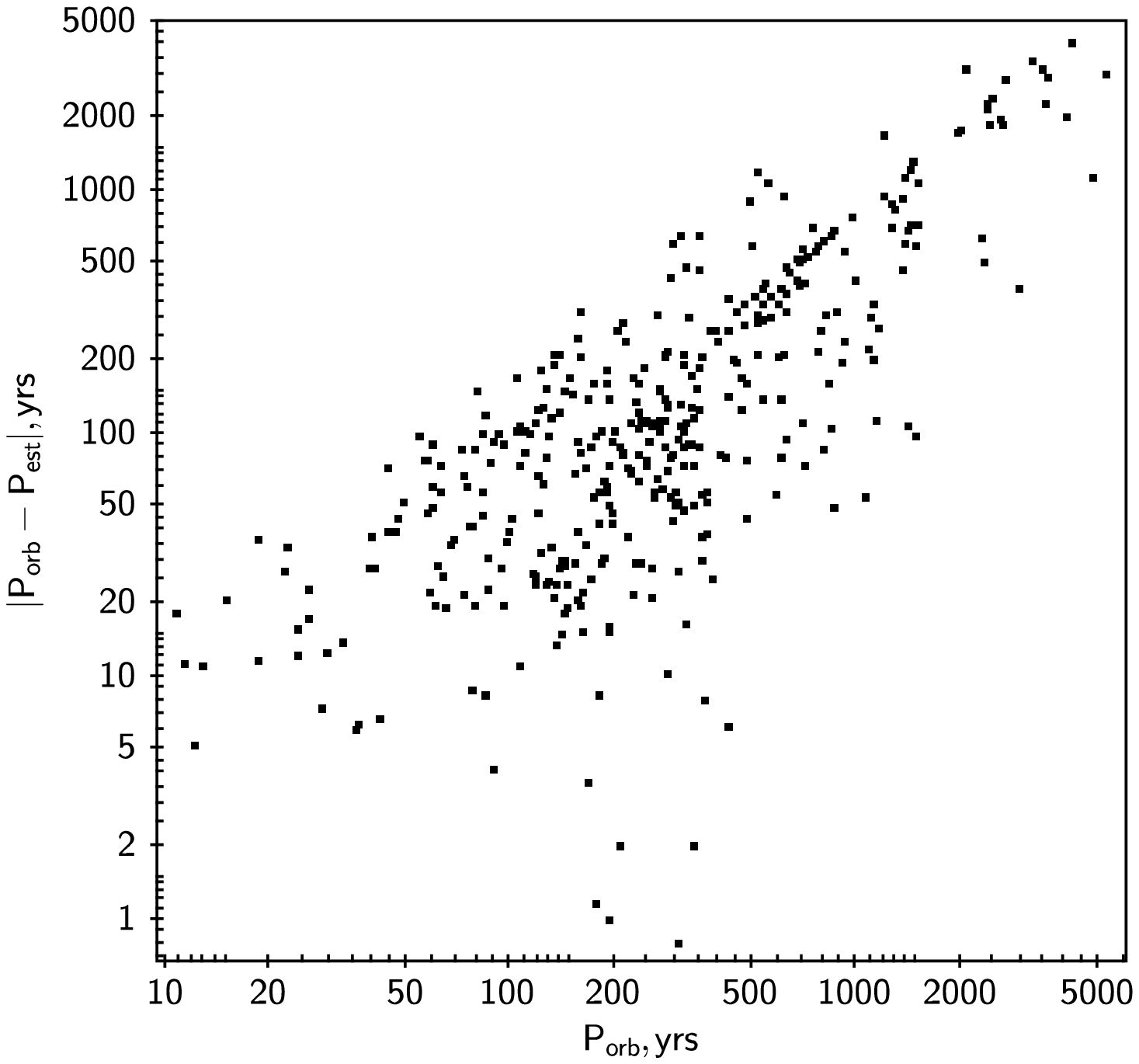,width=100mm,angle=0,clip=}}}
\vspace{1mm} \captionb{2} {The difference between orbital and
estimated period for the same binaries vs orbital period (based on
the data for 833 binaries with orbital periods $\geq$ 10 yrs found
both in the refined WCT dataset and in the ORB6 catalogue).}
\end{figure}

\begin{figure}[!tH]
\vbox{
\centerline{\psfig{figure=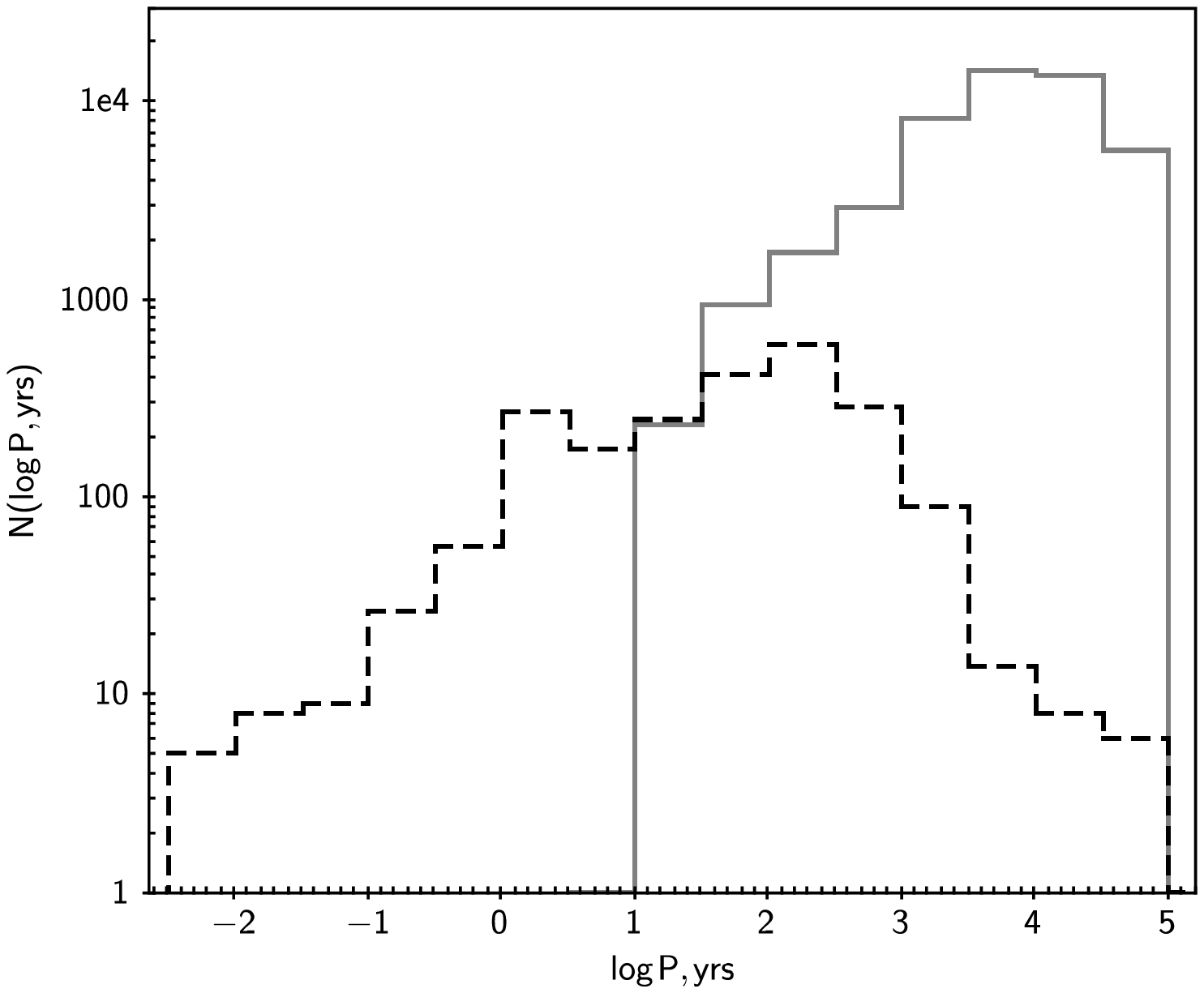,width=100mm,angle=0,clip=}}}
\vspace{1mm} \captionb{3} {Distribution over the period $P$: the
solid line is for the refined WCT dataset and the
dashed line, for all ORB6 periods. }
\end{figure}

\sectionb{2}{DISTRIBUTION OF WIDE BINARIES OVER PERIOD}

The distribution of wide binaries over the period/semimajor axis
of their orbits reflects the process of their formation and can
help to understand the latter. In general, two possibilities are
discussed: the power-law $f(a) \propto  a^{-\alpha}$  (which is,
in the case of $\alpha=1$, the  well-known \"Opik's distribution)
and the Gaussian-like law in respect to $\log P$ or $\log a$.
\"Opik's distribution is an indicator of the energy relaxation
process involving interaction of multiple bodies, which suggests
binary formation preferably or exclusively in multiple systems
(Poveda et al. 2006).

After refining our dataset by removing erroneous data of various
kinds, optical pairs, rounded-up data, multiples, etc. (see
discussion in Kovaleva et al. 2015b), some 47500 pairs having two
and more different valid observations of relative positions of the
components at different epochs remained. For more than 8000 of
them, three or more observations are available. This allows us to
estimate average angular velocity of the visible orbital motion
and thus, to extrapolate it to a rough estimate of the orbital
period. Let us note that, unlike visual binaries with calculated
orbits, the probability for a star to be observed as a visual
binary is related to the period value only for the closest and for
the most remote pairs, while in the wide range of periods it does
not depend on the period. Evaluation of the angular velocity of
visible orbital motion does not require parallax and thus can be
performed for all the 47500 pairs of our dataset.

The internal scatter of such an estimate can be determined for the
same binaries using different pairs of positions where it is
possible. Figure~1 represents the estimates of period for the same
wide pair obtained for two positions in the WDS catalogue vs. one
position in the WDS and one non-matching WDS position from TDSC.
$80\%$ of points are located within $P_2=P_1\pm 50\% $ sector
(grey dots). The orbital periods for almost thousand pairs of our
dataset can be found in the ORB6 catalogue of orbital elements
(Mason et al. 2001). Thus,  we can compare our rough estimates for
these binaries to the period values obtained from orbital
solutions. For this comparison, we excluded primary pairs of
multiple systems to avoid pair identification problems, and pairs
with periods less than 10 years because the method is not
applicable for such short-period pairs. One cannot determine from
two or three observations whether they refer to the same orbital
cycle (as we presume in evaluations). This makes our estimates
systematically larger than orbital periods even for the binaries
with $P \ge 10$ years, but the estimate for shorter periods is
always non-realistic. This limits the dataset down to 833 binary
stars with periods of 10 years or longer.

Figure~2 represents the absolute value of the difference between
the values of orbital periods (obtained from orbital solution) and
periods roughly estimated from angular orbital velocity, vs.
orbital periods. This gives us an estimate of the typical error
range for the method of getting period from the angular orbital
velocity. It appears to be approximately $\pm 40$~years for
periods shorter than 100~years; about $\pm 100$ yrs for $200 \leq
P \leq 500$~years; and about $\pm 600$ yrs for the range of
periods between 1000 and 2000 years.

Figure~3 displays the cumulative distribution over estimated
periods for the pairs of the refined dataset from the WCT. For
comparison, the distribution of all the binaries of ORB6 over the
period is also shown.


\begin{figure}[!tH]
\vbox{
\centerline{\psfig{figure=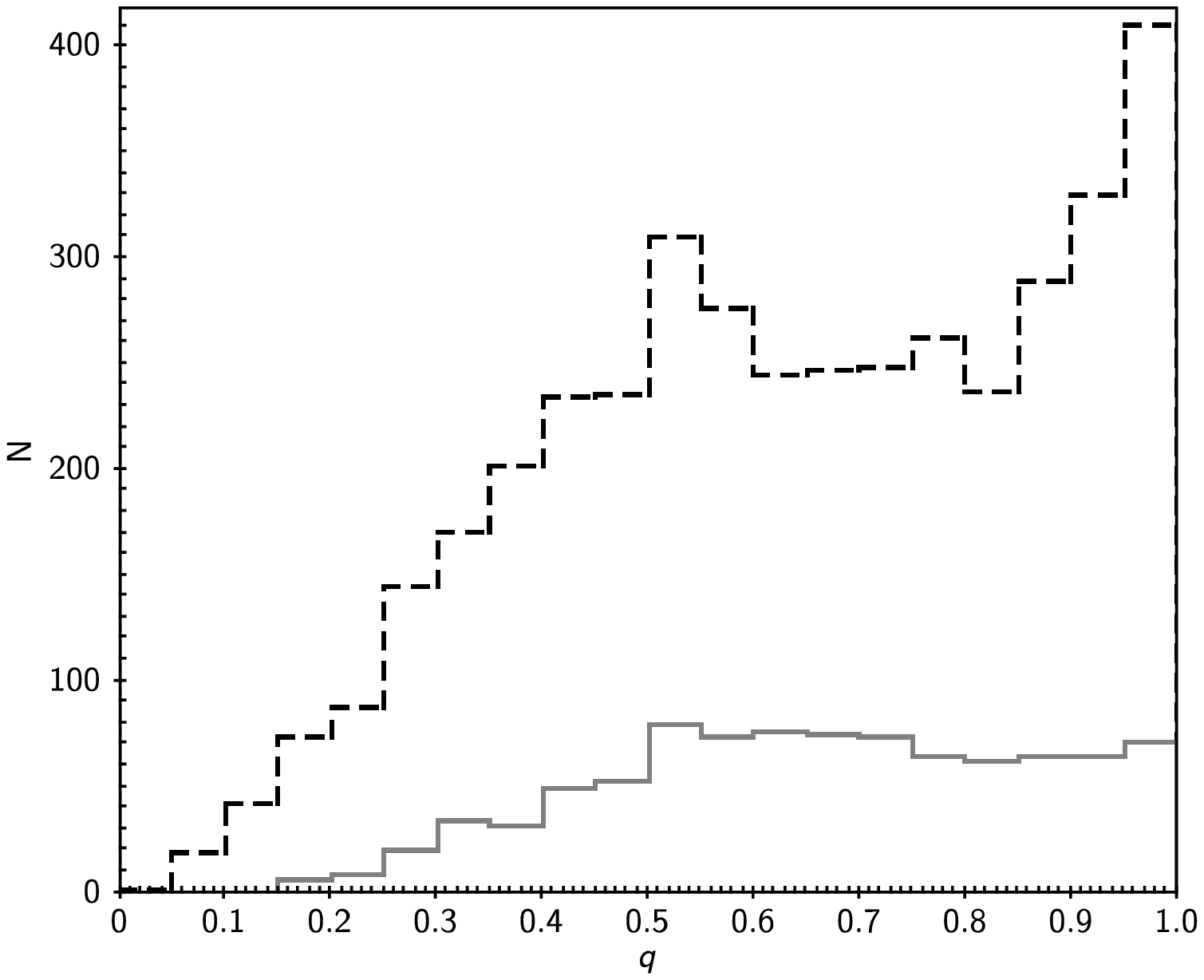,width=100mm,angle=0,clip=}}}
\vspace{1mm} \captionb{4} {Distribution over the mass ratio $q$
obtained for the binaries with main components of luminosity class
V and known trigonometric parallaxes. The dashed line
is for the complete available sample of 4049 binary stars and the
solid line, for the restricted sample of 892 binaries.
The restricted dataset avoids regions of observational
incompleteness in the  $(V_1,\ V_2,\ \Delta V,\ \rho)$ space of
parameters.}
\end{figure}

\sectionb{3}{VERIFICATION OF THE SPECTRAL CLASSIFICATION}

The WCT catalogue  includes spectral classification for about
63000 binaries, while luminosity classes necessary to estimate
physical parameters of the components such as luminosities (if the
distance is unknown) and masses are available for less than a half
of this set. The refined WCT dataset does not include pairs with
degenerate or other peculiar-type components. For other stars,
since mainly only the spectral type of the primary or combined
spectrum is known, we consider this available spectral type as
that of the primary and expect that the secondary has the same or
fainter luminosity class (Kovaleva et al. 2015b).

The investigation of the binaries with primaries' luminosity
class~V in combination with trigonometric parallaxes for these
stars, when available, demonstrated, however, that for this
dataset as a whole (containing 4059 binaries) absolute
luminosities, determined by trigonometric parallax taking into
account interstellar extinction, were systematically larger than
their absolute luminosities prescribed by the luminosity class and
spectral type combination. This suggested that the dataset might
actually contain stars of luminosity classes other than V that
contaminate data statistics, and forced us to reconsider spectral
classifications using alternate sources of data.

We have tested the data on spectral classification within the WCT
catalogue noticing discrepancies between WDS, CCDM, and TDSC data,
and checked the WCT dataset spectra in the SIMBAD database, as
well as in the modern catalogues containing spectral
classification (5268 pairs found in McDonald et al. 2012; 192
pairs identified with the NStars catalogues, Gray et al. 2003,
2006). As the result of this search, we found and corrected
erroneously attributed luminosity classes for 68 of about 1600
giants (luminosity class~III) and for 77 of about 4000 dwarfs
(luminosity class~V) with known trigonometric parallaxes. Thus, we
have compiled a dataset containing 5731 binary stars with
confirmed luminosity class and known trigonometric parallax,
including 4049 pairs with main components having luminosity
class~V and 1682 pairs with main components of class~III.

\sectionb{4}{DETERMINING CHARACTERISTICS OF THE COMPONENTS}

For the set of 4049 pairs with confirmed luminosity class~V and
known trigonometric parallax, we obtain bolometric luminosities
from visible magnitudes, parallaxes, interstellar extinction $A_v$
estimated using the cosecant law (Parenago 1940), and $M_v - \log
L_{\rm bol}$ calibrations for dwarfs from Pecaut \& Mamajek
(2013), Mamajek (2016).  We avoid using multicolor photometry from
SIMBAD or estimates of $\log L_{\rm bol}$ by McDonald et al.
(2012) available for the majority of the sample because it is
mainly uncertain whether it refers to the main component or to the
binary star as a whole.

We determine regions of observational incompleteness in the space
of parameters $(V_1,\ V_2,\ \Delta V,\ \rho)$, as described in
Kovaleva et al. (2015b). The chance for a star to be detected as a
visual binary depends on the visible distance between the
components, magnitude of the primary, and magnitude difference. We
restrict our dataset to $\rho > 1$ arcsec, $V_1 <8.5^m$, $V_2
<10.5^m$, $\Delta V < 4^m$, to avoid regions of the space of
parameters where it is obviously incomplete. This ``restricted''
sample contains 892 binary stars.

Figure~4 presents the mass ratio of the components,  $q=m_2/m_1$,
we obtain for pairs with confirmed luminosity class~V and known
trigonometric parallax. The black dashed broken line refers to the
complete set of 4049 pairs, while the grey solid broken line
represents data for the restricted sample of 892 binary stars. For
this last sample, the mass ratio distribution looks flat for $q
\ge 0.5$.


\sectionb{5}{RESULTS: DATASETS FOR MODELLING FORMATION HISTORY\\ OF WIDE BINARIES}

As the result of cross-matching, analysis, and investigation of
the data for visual binary stars, several datasets have been
compiled for modelling the present-day distributions of wide
binaries over various parameters. For various datasets, the
distributions over orbital periods, semimajor axes, absolute
magnitudes, bolometric luminosities and masses of components were
constructed (Kovaleva et al. 2015b; this paper). The dataset of
892 pairs was corrected for selection effects on stellar
magnitude, magnitude difference and angular separation between the
components. It is still distorted by evolutionary effects and is
not volume-complete.  This dataset allows us to consider some
current distributions of wide binaries on physical characteristics
within a certain range of parameters. The dataset of 5731 pairs
(4049 binary stars with primaries of luminosity classes V and 1682
binaries with primaries of class III) with known physical
characteristics of the components permits to model the observed
distributions of binaries over observational and physical
parameters. The dataset of 47500 pairs, with several reliable
relative positions of the components measured, will be used to
model the distribution over observed medium angular velocity of
orbital motion of the pairs, to discriminate between possible
models of distribution of binaries over orbital periods. The
dataset of about 102000 pairs refined by removing erroneous data,
optical pairs, rounded-up data, multiples, etc. will be used to
develop a statistical error model.

\thanks{The work was partly supported by the Russian Foundation for Basic Research
grants 15-02-04053 and 16-07-01162, Presidium of the Russian
Academy of Sciences Program P-7, and by President Science School
9951.2016.2. This research has made use of the VizieR catalogue
access tool and the SIMBAD database operated at CDS, Strasbourg,
France, the Washington Double Star Catalog maintained at the U.S.
Naval Observatory, and NASA's Astrophysics Data System
Bibliographic Services. }

\References

\refb Abt H. A. 2006, ApJ, 651, 1151

\refb Abt H. A., Levy S. G. 1976, ApJS, 30, 73

\refb Aitken R. G. 1935, ApJ, 82, 368

\refb Ambartsumian V. A. 1937, AZh, 14, 207

\refb Dommanget J., Nys O., 2002, VizieR On-line Data Catalog: I/274

\refb Duch\^ene G., Kraus A., 2013, ARA\&A, 51, 269

\refb Duquennoy A., Mayor M., 1991, A\&Ap, 248, 485

\refb Fabricius C., H{\o}g E., Makarov V. V. et al. 2002, A\&A,
384, 180

\refb Gray R. O., Corbally C. J., Garrison R. F. et al. 2003, AJ,
126, 2048

\refb Gray R. O., Corbally C. J., Garrison R. F. et al. 2006, AJ,
132, 161

\refb Isaeva A. A., Kovaleva D. A., Malkov O. Yu. 2015, Baltic
Astron., 24, 157

\refb Kaygorodov P., Debray B., Kolesnikov N. et al. 2012, Baltic
Astronomy 21, 309

\refb Kovaleva D., Kaygorodov P., Malkov O. et al. 2015a,
Astronomy \& Computing, 11, 119

\refb Kovaleva D., Malkov O., Yungelson L. et al. 2015b, Baltic
Astron., 24, 367

\refb Kouwenhoven M. B. N., Brown A. G. A., Goodwin S. P. et al.
2008, AN, 329, 984

\refb Kraicheva Z. T., Popova E. I., Tutukov A. V., Yungelson L.
R. 1989, Nauchnye Informatsii, 67, 3

\refb Kuiper G. P. 1933, PhD Thesis, Leiden University

\refb Mamajek E., 2016,\\ \url{www.pas.rochester.edu/~emamajek/EEM_dwarf_UBVIJHK_colors_Teff.dat}

\refb Mason B. D., Wycoff G. L., Hartkopf W. I. et al. 2001, AJ,
122, 3466

\refb Mason B.D., Wycoff G.L., Hartkopf W.I., Douglass G.G.,
Worley C.E. 2014, VizieR On-line Data Catalog: B/wds

\refb McDonald I., Zijlstra A. A., Boyer M. L. 2012, MNRAS, 427,
343

\refb \"Opik E. 1924, Tartu Obs. Publ., 25, 6

\refb Patience J., Ghez A. M., Reid I. N., Matthews K. 2002, AJ,
123, 1570

\refb Parenago P. P. 1940, Bull. Sternberg Astron. Inst., 4

\refb Pecaut M. J., Mamajek E. E. 2013, ApJS, 208, 9

\refb Poveda A., Allen C., Hernandez-Alcantara A. 2007, IAU Symp.,
240, 417

\refb Poveda A., Herrera M. A., Allen C. et al. 1994, Rev. Mex.
Astron. Astrophys., 28, 43

\refb Rahgavan D., McAlister H. A., Henry T. J. et al. 2010, ApJS,
190, 1

\refb Tokovinin A. 2014, AJ, 147, 86

\refb Tokovinin A., Kiyaeva O. 2016, MNRAS, 456, 2070

\refb Vereshchagin S., Tutukov A., Yungelson L. et al. 1988, ApSS,
142, 245

\end{document}